\input harvmac
\input epsf
\input amssym
%
%
\noblackbox
\newcount\figno
\figno=0
\def\fig#1#2#3{
\par\begingroup\parindent=0pt\leftskip=1cm\rightskip=1cm\parindent=0pt
\baselineskip=11pt
\global\advance\figno by 1
\midinsert
\epsfxsize=#3
\centerline{\epsfbox{#2}}
\vskip -21pt
{\bf Fig.\ \the\figno: } #1\par
\endinsert\endgroup\par
}
\def\figlabel#1{\xdef#1{\the\figno}}
\def\encadremath#1{\vbox{\hrule\hbox{\vrule\kern8pt\vbox{\kern8pt
\hbox{$\displaystyle #1$}\kern8pt}
\kern8pt\vrule}\hrule}}

\def\frac#1#2{{#1 \over #2}}

\def\p{\partial}
\def\semi{\subset\kern-1em\times\;}

\def\sqr#1#2{{\vcenter{\vbox{\hrule height.#2pt
\hbox{\vrule width.#2pt height#1pt \kern#1pt \vrule width.#2pt}
\hrule height.#2pt}}}}

\def\p{\partial}

\def\bt{{\tilde{b}}}

\def\p{\partial}

\def\dR{^{(d)}\!R}
\def\go{g^{(0)} }
\def\d{^{(d)}\!}

%

%

\def\glt{g_{(2)}}


\lref\fareytale{J.~M.~Maldacena and A.~Strominger,
  ``AdS(3) black holes and a stringy exclusion principle,''
  JHEP {\bf 9812}, 005 (1998)
  [arXiv:hep-th/9804085];
  R.~Dijkgraaf, J.~M.~Maldacena, G.~W.~Moore and E.~Verlinde,
  ``A black hole farey tail,''
  arXiv:hep-th/0005003.

}

\lref\generalRRRR{
  M.~B.~Green and J.~H.~Schwarz,
  ``Supersymmetrical Dual String Theory. 2. Vertices And Trees,''
  Nucl.\ Phys.\ B {\bf 198}, 252 (1982);
  D.~J.~Gross and E.~Witten,
  ``Superstring Modifications Of Einstein's Equations,''
  Nucl.\ Phys.\ B {\bf 277}, 1 (1986);
  W.~Lerche, B.~E.~W.~Nilsson and A.~N.~Schellekens,
  ``Heterotic String Loop Calculation Of The Anomaly Cancelling Term,''
  Nucl.\ Phys.\ B {\bf 289}, 609 (1987);
  M.~J.~Duff, J.~T.~Liu and R.~Minasian,
  ``Eleven-dimensional origin of string / string duality: A one-loop test,''
  Nucl.\ Phys.\ B {\bf 452}, 261 (1995)
  [arXiv:hep-th/9506126];
   M.~B.~Green, M.~Gutperle and P.~Vanhove,
  ``One loop in eleven dimensions,''
  Phys.\ Lett.\ B {\bf 409}, 177 (1997)
  [arXiv:hep-th/9706175];
    J.~G.~Russo and A.~A.~Tseytlin,
  ``One-loop four-graviton amplitude in eleven-dimensional supergravity,''
  Nucl.\ Phys.\ B {\bf 508}, 245 (1997)
  [arXiv:hep-th/9707134]; P.~S.~Howe and D.~Tsimpis,
  ``On higher-order corrections in M theory,''
  JHEP {\bf 0309}, 038 (2003)
  [arXiv:hep-th/0305129].
}

\lref\masakiref{
  N.~Iizuka and M.~Shigemori,
  ``A Note on D1-D5-J System and 5D Small Black Ring,''
  arXiv:hep-th/0506215.
}

\lref\russuss{  J.~G.~Russo and L.~Susskind,
  ``Asymptotic level density in heterotic string theory and rotating black
  holes,''
  Nucl.\ Phys.\ B {\bf 437}, 611 (1995)
  [arXiv:hep-th/9405117].
  }

\lref\KLatt{
  P.~Kraus and F.~Larsen,
  ``Attractors and black rings,''
  Phys.\ Rev.\ D {\bf 72}, 024010 (2005)
  [arXiv:hep-th/0503219].
}

\lref\ChamseddinePI{
  A.~H.~Chamseddine, S.~Ferrara, G.~W.~Gibbons and R.~Kallosh,
  ``Enhancement of supersymmetry near 5d black hole horizon,''
  Phys.\ Rev.\ D {\bf 55}, 3647 (1997)
  [arXiv:hep-th/9610155].
}

\lref\supertube{D.~Mateos and P.~K.~Townsend,
  ``Supertubes,''
  Phys.\ Rev.\ Lett.\  {\bf 87}, 011602 (2001)
  [arXiv:hep-th/0103030];
  R.~Emparan, D.~Mateos and P.~K.~Townsend,
  ``Supergravity supertubes,''
  JHEP {\bf 0107}, 011 (2001)
  [arXiv:hep-th/0106012].
}

\lref\Mathurrev{ S.~D.~Mathur,``The fuzzball proposal for black
holes: An elementary review,'' arXiv:hep-th/0502050.
}

\lref\StromBTZ{ A.~Strominger,
 ``Black hole entropy from near-horizon microstates'',
JHEP {\bf 9802}, 009 (1998); [arXiv:hep-th/9712251];
V.~Balasubramanian and F.~Larsen, ``Near horizon geometry and
black holes in four dimensions'', Nucl.\ Phys.\ B {\bf 528}, 229
(1998); [arXiv:hep-th/9802198].
}

\lref\MSW{ J.~M.~Maldacena, A.~Strominger and E.~Witten, ``Black
hole entropy in M-theory'', JHEP {\bf 9712}, 002 (1997);
[arXiv:hep-th/9711053]. }

\lref\HMM{ J.~A.~Harvey, R.~Minasian and G.~W.~Moore,
``Non-abelian tensor-multiplet anomalies,''
 JHEP {\bf 9809}, 004 (1998)
  [arXiv:hep-th/9808060].
}

\lref\tseytRRRR{  A.~A.~Tseytlin,
  ``R**4 terms in 11 dimensions and conformal anomaly of (2,0) theory,''
  Nucl.\ Phys.\ B {\bf 584}, 233 (2000)
  [arXiv:hep-th/0005072].
}

\lref\antRRRR{  I.~Antoniadis, S.~Ferrara, R.~Minasian and
K.~S.~Narain,
  ``R**4 couplings in M- and type II theories on Calabi-Yau spaces,''
  Nucl.\ Phys.\ B {\bf 507}, 571 (1997)
  [arXiv:hep-th/9707013].
 }

\lref\WittenMfive{ E.~Witten,
  ``Five-brane effective action in M-theory,''
  J.\ Geom.\ Phys.\  {\bf 22}, 103 (1997)
  [arXiv:hep-th/9610234].
}

\lref\wittenAdS{ E.~Witten,
  ``Anti-de Sitter space and holography,''
  Adv.\ Theor.\ Math.\ Phys.\  {\bf 2}, 253 (1998)
  [arXiv:hep-th/9802150].
  }

\lref\iosef{  I.~Bena,
  ``Splitting hairs of the three charge black hole,''
  Phys.\ Rev.\ D {\bf 70}, 105018 (2004)
  [arXiv:hep-th/0404073].
}

\lref\brownhen{  J.~D.~Brown and M.~Henneaux,
 ``Central Charges In The Canonical Realization Of Asymptotic Symmetries: An
  Example From Three-Dimensional Gravity,''
  Commun.\ Math.\ Phys.\  {\bf 104}, 207 (1986).
  }

\lref\wald{
  R.~M.~Wald,
  ``Black hole entropy is the Noether charge,''
  Phys.\ Rev.\ D {\bf 48}, 3427 (1993)
  [arXiv:gr-qc/9307038].
R.~Wald, Phys.\ Rev.\ D {\bf 48} R3427 (1993);
   V.~Iyer and R.~M.~Wald,
  ``Some properties of Noether charge and a proposal for dynamical black hole
  entropy,''
  Phys.\ Rev.\ D {\bf 50}, 846 (1994)
  [arXiv:gr-qc/9403028].
 ``A Comparison of Noether charge and Euclidean methods for computing the
  entropy of stationary black holes,''
  Phys.\ Rev.\ D {\bf 52}, 4430 (1995)
  [arXiv:gr-qc/9503052].
}

\lref\senrescaled{  A.~Sen,
  ``How does a fundamental string stretch its horizon?,''
  JHEP {\bf 0505}, 059 (2005)
  [arXiv:hep-th/0411255];
   ``Black holes, elementary strings and holomorphic anomaly,''
  arXiv:hep-th/0502126;
   ``Stretching the horizon of a higher dimensional small black hole,''
  arXiv:hep-th/0505122; ``Black hole entropy function and the attractor mechanism in higher
  derivative gravity,''
  arXiv:hep-th/0506177; ``Entropy Function for Heterotic Black Holes,''
  arXiv:hep-th/0508042.
  }

\lref\saidasoda{
  H.~Saida and J.~Soda,
  ``Statistical entropy of BTZ black hole in higher curvature gravity,''
  Phys.\ Lett.\ B {\bf 471}, 358 (2000)
  [arXiv:gr-qc/9909061].
}

\lref\attract{ S.~Ferrara, R.~Kallosh and A.~Strominger, ``N=2
extremal black holes'', Phys.\ Rev.\ D {\bf 52}, 5412 (1995),
[arXiv:hep-th/9508072];
 A.~Strominger,
 ``Macroscopic Entropy of $N=2$ Extremal Black Holes'',
 Phys.\ Lett.\ B {\bf 383}, 39 (1996),
[arXiv:hep-th/9602111];
S.~Ferrara and R.~Kallosh, ``Supersymmetry and Attractors'',
Phys.\ Rev.\ D {\bf 54}, 1514 (1996), [arXiv:hep-th/9602136];
``Universality of Supersymmetric Attractors'', Phys.\ Rev.\ D {\bf
54}, 1525 (1996), [arXiv:hep-th/9603090];
R.~Kallosh, A.~Rajaraman and W.~K.~Wong, ``Supersymmetric rotating
black holes and attractors'', Phys.\ Rev.\ D {\bf 55}, 3246
(1997), [arXiv:hep-th/9611094];
A~Chou, R.~Kallosh, J.~Rahmfeld, S.~J.~Rey, M.~Shmakova and
W.~K.~Wong, ``Critical points and phase transitions in 5d
compactifications of M-theory''. Nucl.\ Phys.\ B {\bf 508}, 147
(1997); [arXiv:hep-th/9704142].
}

\lref\moore{G.~W.~Moore,``Attractors and arithmetic'',
[arXiv:hep-th/9807056];
``Arithmetic and attractors'', [arXiv:hep-th/9807087];
``Les Houches lectures on strings and arithmetic'',
[arXiv:hep-th/0401049];
B.~R.~Greene and C.~I.~Lazaroiu, ``Collapsing D-branes in
Calabi-Yau moduli space. I'', Nucl.\ Phys.\ B {\bf 604}, 181
(2001), [arXiv:hep-th/0001025]. }

\lref\ChamseddinePI{
  A.~H.~Chamseddine, S.~Ferrara, G.~W.~Gibbons and R.~Kallosh,
  ``Enhancement of supersymmetry near 5d black hole horizon,''
  Phys.\ Rev.\ D {\bf 55}, 3647 (1997)
  [arXiv:hep-th/9610155].
}

\lref\denef{  
F.~Denef,``Supergravity flows and D-brane stability'', JHEP {\bf
0008}, 050 (2000), [arXiv:hep-th/0005049];
``On the correspondence between D-branes and stationary
supergravity
 solutions of type II Calabi-Yau compactifications'',
[arXiv:hep-th/0010222];
``(Dis)assembling special Lagrangians'', [arXiv:hep-th/0107152].
  B.~Bates and F.~Denef,
   ``Exact solutions for supersymmetric stationary black hole composites,''
  arXiv:hep-th/0304094.
}

\lref\OSV{H.~Ooguri, A.~Strominger and C.~Vafa, ``Black hole
attractors and the topological string'', Phys.\ Rev.\ D {\bf 70},
106007 (2004), [arXiv:hep-th/0405146];
E.~Verlinde, ``Attractors and the holomorphic anomaly'',
[arXiv:hep-th/0412139];
  }

 \lref\DDMP{
A.~Dabholkar, F.~Denef, G.~W.~Moore and B.~Pioline, ``Exact and
asymptotic degeneracies of small black holes'',
[arXiv:hep-th/0502157].
}

\lref\curvcorr{A.~Dabholkar, ``Exact counting of black hole
microstates", [arXiv:hep-th/0409148],
A.~Dabholkar, R.~Kallosh and A.~Maloney, ``A stringy cloak for a
classical singularity'', JHEP {\bf 0412}, 059 (2004),
[arXiv:hep-th/0410076].
} \lref\bkmicro{
 I.~Bena and P.~Kraus,
 ``Microscopic description of black rings in AdS/CFT'',
JHEP {\bf 0412}, 070 (2004)
  [arXiv:hep-th/0408186].
} \lref\cgms{ M.~Cyrier, M.~Guica, D.~Mateos and A.~Strominger,
``Microscopic entropy of the black ring'', [arXiv:hep-th/0411187].
}

\lref\GunaydinBI{
  M.~Gunaydin, G.~Sierra and P.~K.~Townsend,
  ``The Geometry Of N=2 Maxwell-Einstein Supergravity And Jordan
  Algebras'',
  Nucl.\ Phys.\ B {\bf 242}, 244 (1984);
  ``Gauging The D = 5 Maxwell-Einstein Supergravity Theories:
   More On Jordan Algebras,''
  Nucl.\ Phys.\ B {\bf 253}, 573 (1985).
}

\lref\deWitCR{
  B.~de Wit and A.~Van Proeyen,
  ``Broken sigma model isometries in very special geometry,''
  Phys.\ Lett.\ B {\bf 293}, 94 (1992)
  [arXiv:hep-th/9207091].
}

\lref\CadavidBK{
  A.~C.~Cadavid, A.~Ceresole, R.~D'Auria and S.~Ferrara,
  ``Eleven-dimensional supergravity compactified on Calabi-Yau threefolds,''
  Phys.\ Lett.\ B {\bf 357}, 76 (1995)
  [arXiv:hep-th/9506144].
}

\lref\PapadopoulosDA{
  G.~Papadopoulos and P.~K.~Townsend,
  ``Compactification of D = 11 supergravity on spaces of exceptional
  holonomy,''
  Phys.\ Lett.\ B {\bf 357}, 300 (1995)
  [arXiv:hep-th/9506150].
}

\lref\AntoniadisCY{
  I.~Antoniadis, S.~Ferrara and T.~R.~Taylor,
  ``N=2 Heterotic Superstring and its Dual Theory in Five Dimensions,''
  Nucl.\ Phys.\ B {\bf 460}, 489 (1996)
  [arXiv:hep-th/9511108].
}

\lref\GauntlettNW{
  J.~P.~Gauntlett, J.~B.~Gutowski, C.~M.~Hull, S.~Pakis and H.~S.~Reall,
  ``All supersymmetric solutions of minimal supergravity in five dimensions,''
  Class.\ Quant.\ Grav.\  {\bf 20}, 4587 (2003)
  [arXiv:hep-th/0209114]}

\lref\gutow{
  J.~B.~Gutowski and H.~S.~Reall,
  ``General supersymmetric AdS(5) black holes'',
  JHEP {\bf 0404}, 048 (2004)
  [arXiv:hep-th/0401129];
  J.~B.~Gutowski,
  ``Uniqueness of five-dimensional supersymmetric black holes'',
  JHEP {\bf 0408}, 049 (2004)
  [arXiv:hep-th/0404079].
}

\lref\BenaDE{
  I.~Bena and N.~P.~Warner,
  ``One ring to rule them all ... and in the darkness bind them?'',
  [arXiv:hep-th/0408106].
}
\lref\BMPV{ J.~C.~Breckenridge, R.~C.~Myers, A.~W.~Peet and
C.~Vafa, ``D-branes and spinning black holes'', Phys.\ Lett.\ B
{\bf 391}, 93 (1997); [arXiv:hep-th/9602065].
}

\lref\EEMR{H.~Elvang, R.~Emparan, D.~Mateos and H.~S.~Reall,
``Supersymmetric black rings and three-charge supertubes'',
  Phys.\ Rev.\ D {\bf 71}, 024033 (2005);
  [arXiv:hep-th/0408120].
}

\lref\ElvangRT{
  H.~Elvang, R.~Emparan, D.~Mateos and H.~S.~Reall,
  ``A supersymmetric black ring'',
  Phys.\ Rev.\ Lett.\  {\bf 93}, 211302 (2004)
  [arXiv:hep-th/0407065].
}

\lref\BenaWT{
  I.~Bena and P.~Kraus,
  ``Three charge supertubes and black hole hair,''
  Phys.\ Rev.\ D {\bf 70}, 046003 (2004)
  [arXiv:hep-th/0402144].
}

\lref\GauntlettQY{ J.~P.~Gauntlett and J.~B.~Gutowski, ``General
Concentric  Black Rings'', [arXiv:hep-th/0408122]. J.~P.~Gauntlett
and J.~B.~Gutowski, ``Concentric  black rings'',
[arXiv:hep-th/0408010].
}

\lref\denefa{
  F.~Denef,
   ``Supergravity flows and D-brane stability'',
JHEP {\bf 0008}, 050 (2000), [arXiv:hep-th/0005049].
}

\lref\denefc{
  B.~Bates and F.~Denef,
``Exact solutions for supersymmetric stationary black hole
composites'', [arXiv:hep-th/0304094].
}

\lref\SenPU{ A.~Sen, ``Black holes, elementary strings and
holomorphic anomaly'',
 [arXiv:hep-th/0502126].
}

\lref\CardosoFP{
  K.~Behrndt, G.~Lopes Cardoso, B.~de Wit, D.~Lust, T.~Mohaupt and W.~A.~Sabra,
  ``Higher-order black-hole solutions in N = 2 supergravity and Calabi-Yau
  string backgrounds,''
  Phys.\ Lett.\ B {\bf 429}, 289 (1998)
  [arXiv:hep-th/9801081];
G.~L.~Cardoso, B.~de Wit, J.~Kappeli and T.~Mohaupt, ``Examples of
stationary BPS solutions in N = 2 supergravity theories  with
$R^2$-interactions,'' Fortsch.\ Phys.\  {\bf 49}, 557 (2001)
[arXiv:hep-th/0012232]; ``Stationary BPS solutions in N = 2
supergravity with $R^2 $ interactions'', JHEP {\bf 0012}, 019
(2000) [arXiv:hep-th/0009234];
  ``Supersymmetric black hole solutions with $R^2$ interactions'',
[arXiv:hep-th/0003157];
  G.~Lopes Cardoso, B.~de Wit and T.~Mohaupt,
  ``Area law corrections from state counting and supergravity'',
  Class.\ Quant.\ Grav.\  {\bf 17}, 1007 (2000)
  [arXiv:hep-th/9910179];
  ``Macroscopic entropy formulae and non-holomorphic corrections for
  supersymmetric black holes'',
  Nucl.\ Phys.\ B {\bf 567}, 87 (2000)
  [arXiv:hep-th/9906094];
  ``Deviations from the area law for supersymmetric black holes'',
  Fortsch.\ Phys.\  {\bf 48}, 49 (2000)
  [arXiv:hep-th/9904005];
  ``Corrections to macroscopic supersymmetric black-hole entropy'',
  Phys.\ Lett.\ B {\bf 451}, 309 (1999)
  [arXiv:hep-th/9812082].
}

\lref\BenaTD{
  I.~Bena, C.~W.~Wang and N.~P.~Warner,
  ``Black rings with varying charge density'',
[arXiv:hep-th/0411072].
}

\lref\BenaWV{ I.~Bena, ``Splitting hairs of the three charge black
hole'', Phys.\ Rev.\ D {\bf 70}, 105018 (2004),
[arXiv:hep-th/0404073].
}

\lref\rings{ H.~Elvang, R.~Emparan, D.~Mateos and H.~S.~Reall, ``A
supersymmetric black ring,''  Phys. Rev. Lett.  {\bf 93}, 211302
(2004) [arXiv:hep-th/0407065]; ``Supersymmetric black rings and
three-charge supertubes,'' Phys. Rev. D {\bf 71}, 024033 (2005)
[arXiv:hep-th/0408120]; I.~Bena and N.~P.~Warner, ``One ring to
rule them all ... and in the darkness bind them?,''
arXiv:hep-th/0408106; J.~P.~Gauntlett and J.~B.~Gutowski,
``General concentric black rings,'' Phys.\ Rev. D {\bf 71}, 045002
(2005) [arXiv:hep-th/0408122].}

\lref\hensken{  M.~Henningson and K.~Skenderis,
  ``The holographic Weyl anomaly,''
  JHEP {\bf 9807}, 023 (1998)
  [arXiv:hep-th/9806087].
  }

\lref\balkraus{  V.~Balasubramanian and P.~Kraus,
  ``A stress tensor for anti-de Sitter gravity,''
  Commun.\ Math.\ Phys.\  {\bf 208}, 413 (1999)
  [arXiv:hep-th/9902121].
  }

\lref\MarolfFY{
  D.~Marolf and B.~C.~Palmer,
  ``Gyrating strings: A new instability of black strings?,''
  Phys.\ Rev.\ D {\bf 70}, 084045 (2004)
  [arXiv:hep-th/0404139].
}

\lref\MyersPS{
  R.~C.~Myers,
  ``Stress tensors and Casimir energies in the AdS/CFT correspondence,''
  Phys.\ Rev.\ D {\bf 60}, 046002 (1999)
  [arXiv:hep-th/9903203].
}

\lref\KrausDI{
  P.~Kraus, F.~Larsen and R.~Siebelink,
  ``The gravitational action in asymptotically AdS and flat spacetimes,''
  Nucl.\ Phys.\ B {\bf 563}, 259 (1999)
  [arXiv:hep-th/9906127].
}

\lref\deHaroXN{
  S.~de Haro, S.~N.~Solodukhin and K.~Skenderis,
  ``Holographic reconstruction of spacetime and renormalization in the  AdS/CFT
  correspondence,''
  Commun.\ Math.\ Phys.\  {\bf 217}, 595 (2001)
  [arXiv:hep-th/0002230].
}

\lref\PapadimitriouII{
  I.~Papadimitriou and K.~Skenderis,
  ``Thermodynamics of asymptotically locally AdS spacetimes,''
  arXiv:hep-th/0505190.
}

\lref\HollandsWT{
  S.~Hollands, A.~Ishibashi and D.~Marolf,
  ``Comparison between various notions of conserved charges in asymptotically
  AdS-spacetimes,''
  Class.\ Quant.\ Grav.\  {\bf 22}, 2881 (2005)
  [arXiv:hep-th/0503045]; ``Counter-term charges generate bulk symmetries,''
  arXiv:hep-th/0503105.
}

\lref\AlvarezGaumeIG{
  L.~Alvarez-Gaume and E.~Witten,
  ``Gravitational Anomalies,''
  Nucl.\ Phys.\ B {\bf 234}, 269 (1984).
}

\lref\GinspargQN{
  P.~H.~Ginsparg,
  ``Applications Of Topological And Differential Geometric Methods To Anomalies
  In Quantum Field Theory,''
HUTP-85/A056
{\it To appear in Proc. of 16th GIFT Seminar on Theoretical
Physics, Jaca, Spain, Jun 3-7, 1985} }

\lref\us{
  P.~Kraus and F.~Larsen,
  ``Microscopic black hole entropy in theories with higher derivatives,''
  arXiv:hep-th/0506176.
}

\lref\StromingerYG{
  A.~Strominger,
  ``AdS(2) quantum gravity and string theory,''
  JHEP {\bf 9901}, 007 (1999)
  [arXiv:hep-th/9809027].
}

\lref\mtw{C.W. Misner, K.~Thorne, and J.A.~Wheeler,
``Gravitation,"  W. H. Freeman, (1973).}

\lref\BardeenPM{
  W.~A.~Bardeen and B.~Zumino,
  ``Consistent And Covariant Anomalies In Gauge And Gravitational Theories,''
  Nucl.\ Phys.\ B {\bf 244}, 421 (1984).
}

\lref\AlvarezGaumeDR{
  L.~Alvarez-Gaume and P.~H.~Ginsparg,
  ``The Structure Of Gauge And Gravitational Anomalies,''
  Annals Phys.\  {\bf 161}, 423 (1985)
  [Erratum-ibid.\  {\bf 171}, 233 (1986)].
}

\lref\BrownBR{
  J.~D.~Brown and J.~W.~.~York,
  ``Quasilocal energy and conserved charges derived from the gravitational
  Phys.\ Rev.\ D {\bf 47}, 1407 (1993).
}

\lref\fef{C. Fefferman and C.R. Graham, ``Conformal Invariants",
in {\it Elie Cartan et les Math\'{e}matiques d'aujourd'hui}
(Ast\'{e}risque, 1985) 95.}

\lref\btz{
  M.~Banados, C.~Teitelboim and J.~Zanelli,
  ``The Black hole in three-dimensional space-time,''
  Phys.\ Rev.\ Lett.\  {\bf 69}, 1849 (1992)
  [arXiv:hep-th/9204099];  M.~Banados, M.~Henneaux, C.~Teitelboim and J.~Zanelli,
  ``Geometry of the (2+1) black hole,''
  Phys.\ Rev.\ D {\bf 48}, 1506 (1993)
  [arXiv:gr-qc/9302012].
}

\lref\BilalPH{
  A.~Bilal and C.~S.~Chu,
  ``A note on the chiral anomaly in the AdS/CFT correspondence and 1/N**2
  correction,''
  Nucl.\ Phys.\ B {\bf 562}, 181 (1999)
  [arXiv:hep-th/9907106];
  ``Testing the AdS/CFT correspondence beyond large N,''
  arXiv:hep-th/0003129.
}

\lref\MansfieldZW{
  P.~Mansfield and D.~Nolland,
  ``Order 1/N**2 test of the Maldacena conjecture: Cancellation of the
  one-loop Weyl anomaly,''
  Phys.\ Lett.\ B {\bf 495}, 435 (2000)
  [arXiv:hep-th/0005224];
  P.~Mansfield, D.~Nolland and T.~Ueno,
  ``Order 1/N**2 test of the Maldacena conjecture. II: The full bulk  one-loop
  contribution to the boundary Weyl anomaly,''
  Phys.\ Lett.\ B {\bf 565}, 207 (2003)
  [arXiv:hep-th/0208135].
}

\lref\WittenHC{
  E.~Witten,
  ``Five-brane effective action in M-theory,''
  J.\ Geom.\ Phys.\  {\bf 22}, 103 (1997)
  [arXiv:hep-th/9610234].
}

\lref\AharonyRZ{
  O.~Aharony, J.~Pawelczyk, S.~Theisen and S.~Yankielowicz,
  ``A note on anomalies in the AdS/CFT correspondence,''
  Phys.\ Rev.\ D {\bf 60}, 066001 (1999)
  [arXiv:hep-th/9901134].
}

\lref\BlauVZ{
  M.~Blau, K.~S.~Narain and E.~Gava,
  ``On subleading contributions to the AdS/CFT trace anomaly,''
  JHEP {\bf 9909}, 018 (1999)
  [arXiv:hep-th/9904179].
}

\lref\NaculichXU{
  S.~G.~Naculich, H.~J.~Schnitzer and N.~Wyllard,
  ``1/N corrections to anomalies and the AdS/CFT correspondence for
  orientifolded N = 2 orbifold models and N = 1 conifold models,''
  Int.\ J.\ Mod.\ Phys.\ A {\bf 17}, 2567 (2002)
  [arXiv:hep-th/0106020].
}  

\lref\FerraraDD{
  S.~Ferrara and R.~Kallosh,
  ``Supersymmetry and Attractors,''
  Phys.\ Rev.\ D {\bf 54}, 1514 (1996)
  [arXiv:hep-th/9602136].
}

\lref\NojiriMH{
  S.~Nojiri and S.~D.~Odintsov,
  ``On the conformal anomaly from higher derivative gravity in AdS/CFT
  Int.\ J.\ Mod.\ Phys.\ A {\bf 15}, 413 (2000)
  [arXiv:hep-th/9903033].
}

\lref\DeserWH{
  S.~Deser, R.~Jackiw and S.~Templeton,
  ``Topologically Massive Gauge Theories,''
  Annals Phys.\  {\bf 140}, 372 (1982)
  [Erratum-ibid.\  {\bf 185}, 406.1988\ APNYA,281,409
  (1988\ APNYA,281,409-449.2000)]; ``Three-Dimensional Massive Gauge Theories,''
  Phys.\ Rev.\ Lett.\  {\bf 48}, 975 (1982).
}

\lref\JackiwMN{
  R.~Jackiw,
  ``Fifty years of Yang-Mills theory and my contribution to it,''
  arXiv:physics/0403109.
}


\Title{\vbox{\baselineskip12pt
}} {\vbox{\centerline {Holographic Gravitational Anomalies}}}
\centerline{Per
Kraus\foot{pkraus@physics.ucla.edu} and Finn
Larsen\foot{larsenf@umich.edu}}

\bigskip
\centerline{${}^1$\it{Department of Physics and Astronomy,
UCLA,}}\centerline{\it{ Los Angeles, CA 90095-1547,
USA.}}\vskip.2cm \centerline{${}^2$\it{Michigan Center for
Theoretical Physics, Department of Physics}} \centerline{\it{University of Michigan, Ann
Arbor, MI 48109-1120, USA.}}

\baselineskip15pt

\vskip .3in

\centerline{\bf Abstract}

In the AdS/CFT correspondence one encounters theories that are not
invariant under diffeomorphisms.  In the boundary theory this is a
gravitational anomaly, and can arise in $4k+2$ dimensions. In the
bulk, there can be gravitational Chern-Simons terms which vary by
a total derivative.  We work out the holographic stress tensor for
such theories, and demonstrate agreement between the bulk and
boundary. Anomalies lead to novel effects, such as a nonzero
angular momentum for global AdS$_3$.   In string theory such
Chern-Simons terms are known with exact coefficients. The
resulting anomalies, combined with symmetries, imply corrections
to the Bekenstein-Hawking entropy of black holes that agree
exactly with the microscopic counting.

\Date{August, 2005}
\baselineskip14pt
\newsec{Introduction}

Given a theory of gravity in $d+1$ dimensional anti-de Sitter
spacetime, it is possible to define a boundary stress tensor as
the variation of the on-shell action with respect to the metric on
the conformal boundary of the spacetime,
\eqn\aa{ \delta S = \half \int \! d^dx \sqrt{g^{(0)} }\,
T^{ij}\delta g^{(0)}_{ij}~. }
Here $g^{(0)}_{ij}$ is the metric on the conformal boundary, as
reviewed below.  The AdS/CFT correspondence asserts that $T^{ij}$
is the expectation value of the stress tensor in a CFT defined on
a space with metric conformal to $g^{(0)}_{ij}$.  This relation
has been the subject of much work, {\it e.g.}
\refs{\balkraus\MyersPS\KrausDI\deHaroXN\HollandsWT-\PapadimitriouII}.

If the gravitational action is diffeomorphism invariant we will
have $\delta S=0$ for $\delta g^{(0)}_{ij} = \nabla_{(i} \xi
_{j)}$. Inserting this  into \aa\ and integrating by parts we
conclude that the stress tensor is conserved: $\nabla_i T^{ij}
=0$. But this is clearly not the most general situation, since
quantum field theories in $4k+2$ dimensions can have gravitational
anomalies rendering the stress tensor non-conserved
\refs{\AlvarezGaumeIG\BardeenPM-\AlvarezGaumeDR} (see
\refs{\GinspargQN} for a pedagogical review).   For example, this
is the situation for CFTs in two dimensions with unequal left and right
moving central charges. Gravitational anomalies do not spoil the
consistency of a quantum field theory, but they do make it
impossible to consistently couple the theory to dynamical gravity.

The AdS/CFT correspondence thus forces us to confront
non-diffeomorphism invariant theories of gravity in the bulk in
order to account for the non-conservation of the boundary stress
tensor. Such theories will be inconsistent unless the
non-invariance is of a very special type, namely a pure boundary
term.  Variation of the action by a boundary term is harmless
since it does not affect the local dynamics, and is just what we
need to agree with the gravitational anomaly of the boundary
theory.  At the two-derivative order the bulk action is described
by the Einstein-Hilbert term supplemented by boundary terms, and
is diffeomorphism invariant.  Higher derivative terms constructed
covariantly from curvature tensors and matter fields do not change
this conclusion. The only exceptions are Chern-Simons terms
\DeserWH, the purely gravitational version being

\eqn\ab{ S = \beta \int\! \Omega_{d+1}~,}
where the Chern-Simons form is defined as a solution to \eqn\abz{
d\Omega_{d+1} = \Tr~R^{(d+2)/2}~. }
Assuming that we have the  correct coefficient $\beta$ in front of
\ab, variation of the action with respect to a diffeomorphism will
lead to precise agreement with the gravitational anomaly of the
boundary CFT.    This anomaly mechanism is the gravitational analog
of the Chern-Simons gauge anomaly mechanism explained in \wittenAdS.

The purpose of this paper, which is in a sense a continuation of
our previous work \refs{\us}, is to flesh out the details of the
stress tensor in the presence of gravitational Chern-Simons terms.
We will focus  primarily on the simplest case of AdS$_3$. The
anomaly leads to a non-conserved  stress tensor only when the
boundary metric is curved; nevertheless, we will see that  the
anomaly still makes its presence known even for simple metrics
with flat boundary such as global AdS$_3$. In particular, global
AdS$_3$ acquires a nonzero angular momentum,
\eqn\ac{ J = 4\pi \beta~.}
In the boundary CFT this is to be thought
of as a ``Casimir momentum" circulating around the boundary. Indeed,
in the presence of the anomaly $c_L \neq c_R$, and so the left and
right moving zero point momenta do not cancel.  The central charges
are given by
\eqn\aca{ c_L = c_0 + 48\pi \beta,\quad c_R =c_0- 48\pi \beta~,}
where $c_0$ is the central charge in the absence of the Chern-Simons
term.  We also consider rotating BTZ black holes. The geometry
itself is uncorrected by the presence of \ab, but the expressions
for the mass and angular momentum are shifted. Hence the entropy
formula, expressed in terms of the mass and angular momentum, is
also corrected.

There has recently been much interest in the computation of
corrections to the Bekenstein-Hawking area law formula in string
theory due to the presence of higher derivatives
\refs{\MSW\HMM\CardosoFP\curvcorr\senrescaled\DDMP\OSV-
\masakiref,\us}. In certain cases, detailed agreement between
microscopic and macroscopic computations has been exhibited to all
orders in an inverse charge expansion. However, the success of
these comparisons was initially somewhat mysterious, because on
the gravity side technical limitations only allow one to include
the effect of a certain subset of higher derivative terms.  The
puzzle was that the omitted terms individually contribute and so
will spoil the agreement unless there is some cancellation
mechanism.

In \refs{\us}, building on observations in \refs{\HMM}, we showed
that this cancellation mechanism follows from symmetries and
anomalies, and this further led to a much simplified derivation of
the nonzero corrections to the entropy. In particular, to derive
the corrections all one needs to know is the coefficient of a
particular Chern-Simons term in the action, and the exact value of
this coefficient is easily determined from anomalies.  Here we
will give some more details regarding this analysis.

Most work on the subject of higher derivative corrections has been
in  the context of black holes with geometry AdS$_2 \times S^2
\times X$. However, the AdS$_2$ factor is the ``very near horizon"
limit of an AdS$_3$ factor \refs{\StromingerYG}, and so one can
instead work with AdS$_3 \times S^2 \times Y$.  The latter
representation is preferable for our purposes since it makes the
full conformal symmetry manifest. In this geometry we can consider
anomalies associated with diffeomorphisms in AdS$_3$ or on $S^2$.
The former gives the gravitational anomaly of the CFT, and so
determines $c_L - c_R$. The latter is interpreted as the $SU(2)$
R-symmetry anomaly in the CFT; by supersymmetry this fixes $c_L$. We
can therefore determine $c_L$ and $c_R$ exactly from anomalies \HMM.
As explained in \refs{\us}, knowledge of the central charges is
sufficient to derive the corrections to the entropy, even when
higher derivative corrections are taken into account. The result
reproduces the higher derivative entropy formulas in the literature,
and explains why they are correct.  We also extended the class of
examples to include non-BPS states, and states with nonzero angular
momentum. Again, it is the powerful constraint of symmetries and
anomalies that allow us to make exact statements in these cases.

Gravitational anomalies can show up either in diffeomorphisms,
rendering the stress tensor non-conserved, or, if one adopts the
vielbein formalism,  in local Lorentz transformations, rendering
the stress tensor non-symmetric.  These are equivalent in the
sense that there exists a counterterm that can be added to the
action to shift the anomaly from one form to the other
\refs{\BardeenPM,\AlvarezGaumeDR}. One outcome of our analysis is
a simple expression for this counterterm in AdS$_3$.

In studying anomalies in quantum field theory it is often
convenient to think of spacetime as being the boundary of a higher
dimensional disk.  Various expressions take a simpler form when
expressed as integrals over the disk.  It is amusing to note that
this can be thought of as an indication of holography, for in this
context the disk is nothing else than Euclidean  AdS$_{d+1}$.

This paper is organized as follows. In section 2 we review the
holographic stress tensor for Einstein gravity. In section 3 we
discuss the nature of the variational principle when higher
derivatives are present, and we introduce basic properties of the
Chern-Simons term. In section 4 we discuss the effect of anomalies
on the stress tensor and on the central charges. In section 5 we
compute the stress tensor explicitly, and apply the result to
global AdS$_3$ as well as the BTZ black hole. Finally, in section
6, we discuss how anomalies determine higher derivative
corrections to the black hole entropy. Some needed technical
results are found in the appendices.

\newsec{Holographic stress tensor in Einstein gravity}

In this section we review the derivation of the holographic stress
tensor in the case of two-derivative Einstein gravity.  This review
will also serve to fix conventions and notation.  Our
curvature conventions follows those in \mtw.

\subsec{The Brown-York Stress Tensor} We work in $D=d+1$ Euclidean
dimensions.  It is convenient to adopt Gaussian normal coordinates
by foliating the spacetime with $d$ dimensional hypersurfaces
labelled by $\eta$, and writing the metric as
\eqn\ba{ds^2 = d\eta^2 + g_{ij} dx^i dx^j~.}
In these coordinates the extrinsic curvature of a fixed $\eta$
surface reads
\eqn\bb{K_{ij} = \half \p_\eta g_{ij}~.}
The $D$ dimensional Ricci scalar decomposes as
\eqn\bc{ R=^{(d)}\!\!R - (\Tr K)^2-\Tr K^2 - 2\p_\eta \Tr K~,}
where $\Tr K = g^{ij} K_{ij}$, and similarly for $\Tr K^2$, and
$\dR$ denotes the Ricci scalar of the metric  $g_{ij}$.

The $D$ dimensional Einstein-Hilbert action is
\eqn\bd{\eqalign{S_{EH}&=  {1 \over 16\pi G } \int_{\cal M} \! d^D
x \sqrt{g} \,(R-2 \Lambda) \cr &= {1 \over 16\pi G } \int_{\cal M}
\! d^dx \,d\eta \sqrt{g} \left(\dR+ (\Tr K)^2- \Tr K^2-2\Lambda
\right) - {1 \over 8 \pi G}\int_{\p {\cal M}} \! d^d x \sqrt{g}\,
\Tr K~,}}
where we have taken the boundary $\partial {\cal M}$ to be a fixed $\eta$ surface. The
variation of the boundary term contains a contribution $\delta
\p_\eta g_{ij}$ whose presence would spoil the variational
principle leading to Einstein's equations.  This is rectified by
adding to the action the Gibbons-Hawking term
\eqn\be{S_{GH} = {1 \over 8 \pi G}\int_{\p {\cal M}} \! d^d x
\sqrt{g}\, \Tr K~. }

We now consider the variation of the action with respect to
$g_{ij}$.  The variation will consist of two terms: a bulk piece
that vanishes when the equations of motion are satisfied, and a
boundary piece.  Assuming that the equations of motion are
satisfied, a simple computation gives
\eqn\bff{\delta (S_{EH} +S_{GH}) = -{1 \over 16\pi G} \int_{\p
{\cal M}}\! d^dx \sqrt{g} \, (K^{ij} - \Tr K g^{ij})\delta
g_{ij}~.}
The stress tensor is defined in terms of the variation as
\eqn\bfa{ \delta S = \half \int_{\p{\cal M}} \! d^dx \sqrt{g}\,
T^{ij} \delta g_{ij}~,}
 and so we have
\eqn\bg{ T^{ij} = -{1 \over 8\pi G} (K^{ij} - \Tr K g^{ij})~,}
which is the result derived by Brown and York \BrownBR. Although
we derived this result in the coordinate system \ba, the result
\bg\ is valid in any coordinate system, where $g_{ij}$ is the
induced metric on the boundary, and $K_{ij}$ is the extrinsic
curvature.

\subsec{Asymptotically anti-de Sitter space}

With a negative cosmological constant,
\eqn\bh{ \Lambda = -{d(d-1) \over \ell^2}~, }
solutions to Einstein's equations admit the expansion
\fef\foot{For AdS$_{2n+1}$ with $n\geq 2$ there is also a term
linear in $\eta$ which is related to the conformal anomaly
\refs{\hensken,\deHaroXN}.  For the remainder of this paper we
will focus on AdS$_3$ where this term is absent \refs{\deHaroXN},
and so we neglect it henceforth.  }
\eqn\bi{ g_{ij}  = e^{2\eta/\ell}g^{(0)}_{ij} + g^{(2)}_{ij} +
e^{-2\eta/\ell} g^{(4)}_{ij}+ \ldots~. }
Such a solution defines a notion of an asymptotically anti-de Sitter
spacetime.  We think of the boundary as being at $\eta= \infty$,
with metric conformal to  $\go_{ij}$.  We then define the stress
tensor in terms of the variation of the action with respect to
$\go_{ij}$ as in \aa.

As explained in
\refs{\wittenAdS,\hensken,\balkraus\MyersPS\KrausDI\deHaroXN\HollandsWT-\PapadimitriouII},
the action and the stress tensor will
diverge unless we include additional counterterms
that are intrinsic to the boundary.  In the case of AdS$_3$ the
counterterm is just the boundary cosmological constant
\eqn\bia{
S_{\rm ct} = -{1\over 8\pi G\ell} \int_{\partial {\cal M}} d^2 x \sqrt{g}~.
}
After including the variation of the counterterm, the result for
the AdS$_3$ stress tensor reads
\eqn\bj{T_{ij} = {1 \over 8\pi G \ell}
\left(g^{(2)}_{ij}-g_{(0)}^{kl}g^{(2)}_{kl} g^{(0)}_{ij}\right)
~.}
Indices are lowered and raised with $\go_{ij}$ and its inverse
$g_{(0)}^{ij}$. The stress tensors for higher dimensional
spacetimes can be found in the references.

\newsec{General aspects of higher derivative theories}
In this section we discuss the addition of higher derivative terms
to the action, either built out of curvature invariants, or as
Chern-Simons terms.

\subsec{The variational principle for higher derivative actions}
The variation of the action with respect to the metric generally
produces a boundary term, as illustrated for Einstein gravity
after \bd.  For an action containing up to $n$ derivatives, the
boundary term will typically involve $(\p_\eta)^k \delta g_{ij}$
with $k=0,1,\ldots n-1$. The standard variational principle seeks
an extremum among configurations with fixed boundary values of the
field, {\it i.e.} it takes $\delta g_{ij}=0$. The terms with
$k\geq 0$ show that solutions to the bulk equations of motion fail
to extremize the action, and so they undercut the standard
variational principle. In a generic higher derivative theory one
might therefore modify the variational principle by imposing
additional boundary conditions, such as $(\p_\eta)^k \delta
g_{ij}=0$.   However, this approach clashes with the AdS/CFT
correspondence, since it would imply the existence of $n-1$
additional ``stress tensors" with no obvious analog in the CFT.
Instead, AdS/CFT implies that we should work with the unmodified
variational principle; there are two ways in which its validity
can be restored.

The first possibility is that there might exist a generalized
Gibbons-Hawking term whose variation precisely cancels the
$(\p_\eta)^k \delta g_{ij}$ terms for $k\geq 1$.   This is a
stringent requirement which is met only for very special choices
of bulk actions.   For instance, given a general four-derivative
action constructed from a linear combination of $R^2$,
 $R^{\mu\nu}R_{\mu\nu}$, and
 $R^{\mu\nu\alpha\beta}R_{\mu\nu\alpha\beta}$, the existence of a
suitable Gibbons-Hawking term fixes the relative coefficients to
be those of the dimensionally continued $D=4$ Euler invariant
\eqn\euler{ S_{\rm Euler} = c_E\left( R^2 - 4R_{\alpha\beta}
R^{\alpha\beta}+ R_{\alpha\beta\gamma\delta}
R^{\alpha\beta\gamma\delta} \right)~. } Generically no generalized
Gibbons-Hawking term exists and so the variational principle
requires the specification of all $(\p_\eta)^k \delta g_{ij}$ on
the boundary, it does not allow them to fluctuate.

The second possibility is to exploit the fact that the boundary is
a surface at infinity, and define falloff conditions such that the
coefficients of the unwanted terms $(\p_\eta)^k \delta g_{ij}$ for
$k\geq 1$ vanish at infinity.  This is what happens for the higher
derivative Chern-Simons terms we consider here, using the
Fefferman-Graham expansion \bi. In terms of \bi, the actual
statement we will need is that the variation of the action only
involves $\delta \go_{ij}$ and not $\delta g^{(2n)}_{ij}$ for
$n>0$.
A natural question, which we do not address here, is to what extent the form
of more complicated higher derivative bulk actions is constrained
by imposing this condition.

\subsec{Gravitational Chern-Simons terms} We now turn our
attention specifically to Chern-Simons terms and their associated
anomalies. Our conventions will follow those in \GinspargQN, which
is  a helpful reference for what follows; see also \JackiwMN.

We define the connection 1-form as
\eqn\bk{\Gamma^\alpha_{~\beta} =\Gamma^\alpha_{\beta \mu}
dx^\mu~,}
where $\Gamma^\alpha_{\beta \mu}$ are the usual Christoffel symbols.
The standard definition of the Riemann tensor is then equivalent to
defining the curvature 2-form as
\eqn\bl{ R^\alpha_{~\beta} = d\Gamma^\alpha_{~\beta} +
\Gamma^\alpha_{~\gamma} \wedge \Gamma^\gamma_{~\beta}~,}
or, using a matrix notation, as
\eqn\bm{ R = d\Gamma + \Gamma \wedge \Gamma~.}
Upon writing $R^\alpha_{~\beta} = \half R^\alpha_{\beta \mu \nu}
dx^\mu \wedge dx^\nu$ we recover the standard component definition
of the Riemann tensor.  To further simplify notation, we will
often suppress the explicit wedge products among forms.

Under the infinitesimal  diffeomorphism $x^\mu \rightarrow x'^\mu=
x^\mu - \xi^\mu(x)$ the metric, connection, and curvature
transform as\foot{To keep formulae simple and to  conform with
\GinspargQN\ we do not shift the argument of the functions
explicitly. Doing so would in any case yield the same variation of
the action as long  as the location of the boundary is kept fixed.
The full change, needed later, amounts to defining
$v^\alpha_{~\beta}$ using a covariant derivative.}
\eqn\bn{ \eqalign{ \delta_\xi g_{\mu\nu} &=v_{\mu\nu} + v_{\nu\mu}~, \cr
\delta_\xi  \Gamma & = dv+[\Gamma,v]~,\cr \delta_\xi R & = [R,v]~, }}
where
\eqn\bo{ v^\alpha_{~\beta} = {\p \xi^\alpha \over \p x^\beta}~.}

The Chern-Simons 3-form
\eqn\bp{ \Omega_3(\Gamma) = \Tr ( \Gamma d\Gamma + {2 \over 3}
\Gamma^3)~}
is central to our applications. It has two key properties: first,
its exterior derivative is a symmetric polynomial
\eqn\br{ d\Omega_3(\Gamma) = \Tr ( R^2)~,}
and second, under a diffeomorphism it varies by a total derivative
\eqn\bq{ \delta_\xi \Omega_3(\Gamma) =
d \Tr (v d\Gamma)~.}
Due to the latter property a term in the action of the form
\eqn\bs{ S_{CS}(\Gamma) =  \int_{{\cal M}} \! \Omega_3(\Gamma)~}
transforms under diffeomorphism by a boundary term,
\eqn\bt{\delta_\xi S_{CS}(\Gamma) = \int_{\p {\cal M}} \! \Tr (v
d\Gamma)~.}
Thus, diffeomorphism invariance of the bulk theory is preserved by
\bs\ in the sense that the equations of motion remain covariant,
even though the full action is not invariant. The holographic
interpretation will be that the dual theory on the boundary
suffers a gravitational anomaly.

It is instructive to consider also an alternative formalism that
preserves diffeomorphism invariance at the expense of local
Lorentz invariance. Now the geometry is represented by the
vielbein $e^a =e^a_{~\mu} dx^\mu$ and, rather than the connection
one-form \bk, we introduce the spin-connection $\omega^a_{~b} =
\omega^a_{~b\mu}dx^\mu$ determined by Cartan's structure equation
\eqn\stru{ de^a  + \omega^a_{~b} e^b =0~. } The curvature 2-form
is then
\eqn\bu{ R^a_{~b}  = d\omega^a_{~b} + \omega^a_{~c}\omega^c_{~b}~.}
Under an infinitesimal local Lorentz transformation parameterized by the matrix
$\Theta^a_{~b}= - \Theta_{b}^{~a}$ we have
\eqn\bv{\eqalign{ \delta_\Theta e & = - \Theta  e~,  \cr
\delta_\Theta \omega &=d\Theta + [\omega,\Theta]~, \cr
\delta_\Theta R & = [R,\Theta]~,}}
where Lorentz indices are implied. Since $e$, $\omega$, and $R$ are differential forms
they are invariant under general coordinate transformations.

In the vielbein formalism we define the Chern-Simons 3-form
\eqn\bw{ \Omega_3(\omega) = \Tr (\omega d\omega +{2 \over 3}
\omega^3)~,}
obeying
\eqn\bx{\eqalign{
 d\Omega_3(\omega) &= \Tr (R^2)~,\cr
 \delta \Omega_3(\omega) &= d\Tr (\Theta d\omega)~.
 }}
Again, a Chern-Simons term in the action,
\eqn\by{S_{CS}(\omega ) =  \int_{{\cal M}} \! \Omega_3(\omega)~,}
has a variation that localizes on the boundary
\eqn\bz{\delta S_{CS}(\omega) = \int_{\p {\cal M}} \! \Tr (\Theta
d\omega)~.}
Thus the bulk theory remains Lorentz invariant in the presence of
the term \by, but the dual boundary theory does not. This is the
alternative manifestation of the gravitational anomaly that we
wanted to exhibit.

For tensors, the transcription between the two alternate
formalisms is the obvious one. For example, the curvature two
forms \bl\ and \bu\  are related by $R^\alpha_{~\beta} =
e^\alpha_{~a}R^a_{~b}e^b_{~\beta}$. However, the relation between
connection one-forms and the spin connections  contains an
inhomogeneous term
\eqn\ca{\Gamma^\alpha_{~\beta} =  e^\alpha_{~a} \omega^a_{~b}
e^b_{~\beta} + e^\alpha_{~a} \p_\mu e^a_{~\beta} dx^\mu~.}
This explains how the  two alternate forms of the Chern-Simons
action \bs\ and \by\ manage to preserve different symmetries: they
are not equal. It is convenient to write \ca\ in a more condensed
fashion as
\eqn\cb{ \Gamma = e^{-1} \omega e + e^{-1} de~.}
Here and in the next three formulas $e$ is interpreted as a matrix
valued 0-form, {\it not} as a 1-form as it is elsewhere; we hope
this will not cause confusion. In this notation, the difference
between the  two forms of the Chern-Simons action is
\eqn\cc{ \Delta S_{CS} \equiv S_{CS}(\Gamma) - S_{CS}(\omega) = -{1
\over 3} \int_{{\cal M}} \! \Tr( e^{-1} de)^3  + \int_{\p {\cal M}}
\!\Tr (\omega de e^{-1})~.}
By construction, $\Delta S_{CS}$ transforms under general coordinate
and local Lorentz transformations as
\eqn\cd{\eqalign{ \delta_\xi \Delta S_{CS} &= \int_{\p {\cal M}} \! \Tr (v
d\Gamma)~,\cr \delta_\Theta \Delta S_{CS} &= -\int_{\p {\cal M}} \!
\Tr (\Theta d\omega)~.}}
Thus, the addition (or subtraction) of $\Delta S_{CS}$ to the
action transforms between the two different forms of the anomaly.
Since the general variation of $\Delta S_{CS}$ localizes on the
boundary
\eqn\ce{ \delta \Delta S_{CS} = -\int_{\p {\cal M}} \! \Tr (\delta
e e^{-1} de e^{-1} de e^{-1}) +\int_{\p {\cal M}} \! \delta \Tr
(\omega de e^{-1})~,}
this term can be interpreted as intrinsic to the boundary  theory.
This means that the two forms of the gravitational anomaly are
equivalent.

\newsec{Holographic gravitational anomalies}
In this section we discuss the effects of gravitational  anomalies
on the boundary stress tensor, and the interpretation of these as
shifts in the central charges.

\subsec{Anomalous Conservation Laws} Conventionally, a general
three dimensional action for gravity coupled to matter is
constructed by forming invariant  terms from curvature tensors,
matter fields, and their covariant derivatives\foot{In general,
additional boundary terms are needed for a well-defined
variational principle, as discussed in section 3.1. Furthermore,
as mentioned in section 2.2, local counter-terms on the boundary
are needed to render the action finite.}. Classically, such an
action would be invariant under general coordinate
transformations\foot{For a manifold with boundary we should demand
that the coordinate transformation does not shift the location of
the boundary.} and local Lorentz transformations. When a
Chern-Simons term is added, these symmetries may be violated by a
nonzero boundary variation as was determined explicitly in the
previous section. This manifests itself in unusual properties of
the boundary stress tensor which we discuss in the following.

We first consider the diffeomorphism anomaly, {\it i.e.} we add
the term $\beta S_{CS}(\Gamma)$ to the action. Quite generally, by
the definition of the stress tensor, we can write the variation of
an action due to a general coordinate transformation as
\eqn\da{\delta_\xi S = \half \int_{\p {\cal M}} \! d^2x \sqrt{g} ~T^{ij}
\delta g_{ij}= \int_{\p {\cal M}} \! d^2x \sqrt{g}~ T^{ij} \nabla_i
\xi_j = - \int_{\p {\cal M}} \! d^2x \sqrt{g}~ \nabla_i T^{ij}
\xi_j~.}
Since the anomaly arises exclusively from the Chern-Simons term we
can compare this with \bt\ and find the anomalous divergence of
the stress tensor
\eqn\db{\nabla_i T^{ij} = g^{ij} \epsilon^{kl} \p_k \p_m
\Gamma^m_{il}~.}

Next, we consider the Lorentz anomaly, {\it i.e.} we add $\beta
S_{CS}(\omega)$ to the action. In the vielbein formalism the
variation of the action is
\eqn\db{\delta_\Theta S =  \int_{\p {\cal M}} \! d^2x~e ~\delta
e^a_{~\mu} T_{ab} e^{b\mu} =- \int_{\p {\cal M}} \! d^2x ~e ~\Theta^{ab}T_{ab}~.}
Comparing this with the variation \bz\ we learn that the stress
tensor picks up an anti- symmetric contribution:
\eqn\dc{T_{ab} - T_{ba}  = 2 \beta~ ^\star\!R_{ab}~,}
where $^\star\!R_{ab}$ is the Hodge dual of the boundary curvature
2-form, {\it i.e.} a 0-form.

Anomalies in general coordinate transformations thus manifest
themselves in non-conservation of the stress tensor, while
anomalies in local Lorentz transformations show up in the
asymmetric part of the stress tensor.  These are not really
independent anomalies since we exhibited in \cc\ a term that can
be added to the action with an appropriate coefficient so as to
cancel one or the other of the anomalies.  Therefore, the
invariant statement is that there is an anomaly in {\it either}
general coordinate or local Lorentz transformations.  We further
remark that this anomaly shifting counterterm $\Delta S_{CS}$ can
be thought of as being defined on the boundary, since its
variation is strictly localized there.  All of this was known from
the early days of gravitational anomalies
\refs{\BardeenPM,\AlvarezGaumeDR}, although the completely
explicit form of the anomaly shifting counterterm was not written
down, as far as we are aware.

\subsec{Anti-de Sitter spacetime}

The results of the last subsection apply to any three dimensional
geometry, including AdS$_3$.  As we have noted, in AdS$_3$ one takes
the conformal boundary metric to be $g^{(0)}_{ij}$.   Our previous
results for the variation of the action apply with the boundary
metric taken to be $g^{(0)}_{ij}$.

In the context of the AdS/CFT correspondence we can compare our bulk
variations to the anomalous variations of the boundary CFT.  If the
boundary theory has central charges $c_L$ and $c_R$ then under a
general coordinate transformation
\eqn\dd{ \delta_\xi S ={c_L -c_R \over 96 \pi} \int_{\p{\cal M}} \Tr(v
d\Gamma)~.}
This assumes we choose to set the local Lorentz anomaly to zero.
Equivalently, if we choose to set the general coordinate anomaly to
zero, then the local Lorentz anomaly is
\eqn\dd{ \delta_\Theta S ={c_L -c_R \over 96 \pi} \int_{\p{\cal M}}
\Tr(\Theta d\omega)~.}
In either case, we learn that a bulk action with Chern-Simons term
$\beta S_{CS}$ corresponds to a theory with
\eqn\de{c_L -c_R = 96 \pi \beta~.}
The Chern-Simons term is  maximally chiral, {\it i.e.} it treats
left and right oppositely. Therefore, the shifts in left and right
central charges must be equal in magnitude, but of opposite sign.
So we can write \de\ as \eqn\dea{ c_L = c_0 + 48\pi \beta~,~~~c_R
= c_0 -48\pi \beta~,} where $c_0$ is the central charge in the
absence of the Chern-Simons term.

\subsec{Gravitational anomaly for AdS$_3 \times S^p$}

Now consider a theory admitting AdS$_3 \times S^p$ as a solution,
with the sphere supported by $p$-form flux
\eqn\df{ -{1 \over 2\pi} \int_{S^p} F^{(p)} = q~.}
We can consider gravitational anomalies associated with
transformations on the sphere.

For definiteness, we phrase the
anomaly in terms of local Lorentz transformations.  Consider the
following deformation of AdS$_3 \times S^p$
\eqn\dg{\eqalign{ ds^2 &= ds_3^2(x) + \sum_{m=1}^p (e^m)^2~,\cr e^m
&= dy^m - A^m_{~n}(x)y^n~, \cr \sum_{m=1}^p (y^m)^2 &= R_{S^p}^2~,}}
where $x^\mu$ and $y^m$ denote the AdS$_3$ and $S^p$ coordinates,
respectively.  $A^m_{~n}$ can be identified with the spin connection
on the sphere: $A^m_{~n}=\omega^m_{~n}$.

Suppose that in this theory there exists a Chern-Simons term of the
form
\eqn\dh{ S_{CS} = \gamma \int\! F^{(p)} \wedge \Omega_3~.}
We can reduce this term to $D=3$ by integrating over $S^p$, yielding
\eqn\di{S_{CS} = \beta  \int\! \Omega_3(\omega) +\beta \int\!
\Omega_3(A)~,}
with $\beta = -2\pi \gamma q$.   We have assumed that $F^{(p)}$
only has components on $S^p$, and so in \di\ we only get the
contribution of the spin connection when its 1-form index is in
AdS$_3$.

The first term in \di\ is the  gravitational Chern-Simons term
that we have studied in the previous sections. The second term is
best interpreted as an $SO(p+1)$ Yang-Mills Chern-Simons term,
where the $SO(p+1)$ gauge invariance corresponds to isometries of
the sphere. The presence of the Chern-Simons term means that there
is a nonzero anomalous boundary contribution associated with
$SO(p+1)$ gauge transformations, $\delta A = d\Lambda
+[A,\Lambda]$,
\eqn\dj{\delta S = \beta \int_{\p {\cal M}} \Tr ( \Lambda dA)~.}

In the context of AdS/CFT, \dj\ is interpreted in the CFT as a
contribution to the R-symmetry anomaly.  Well known cases are the
D1-D5 system, described by $p=3$ and corresponding $SO(4) \cong
SU(2) \times SU(2)$ R-symmetry, and M-theory on CY$_3$ with
wrapped M5-branes, corresponding to $p=2$ and $SU(2)$ R-symmetry.
The latter example is especially relevant in the context of higher
derivative corrections, as we discuss later.

In contexts where AdS$_3$ arises as the decoupled geometry near
some branes, transformations on the sphere amount to  rotations of
the vectors  normal to the brane worldvolume. In this case $A$ is
interpreted as the connection of the normal bundle, and the
associated gravitational anomaly is the normal bundle anomaly
discussed in \WittenHC.

We should emphasize that the  above Yang-Mills Chern-Simons term
arising from \dh\ is a {\it correction}; there is typically such a
term present even starting from the two-derivative Einstein
action, although its derivation can be somewhat subtle \HMM.
Finally, we remark that corrections to anomalies in the AdS$_5$
context have been discussed in
\refs{\AharonyRZ\NojiriMH\BlauVZ\BilalPH\MansfieldZW-\NaculichXU}

\newsec{Holographic stress tensor in the presence
 of Chern-Simons terms}

In this section we compute the contribution  of the gravitational
Chern-Simons term to the boundary stress tensor in an
asymptotically AdS$_3$ spacetime and apply the result to the case
where the bulk geometry is either global AdS$_3$ or the BTZ black
hole.

\subsec{Derivation}

We want to find the contribution of
\eqn\ea{ S_{CS}(\Gamma) = \int_{\cal M} \! \Omega_3(\Gamma)
=\int_{\cal M} \! \left( \Gamma d\Gamma +{2 \over 3}\Gamma^3
\right)~~, }
to the stress tensor. To do this we must work out the change in \ea\
due to a variation of the metric around a solution of the equations
of motion. A useful first step is to vary the connection. This yields
\eqn\ed{ \delta S_{CS}(\Gamma) = 2 \int_{\cal M}\! \Tr( \delta
\Gamma \wedge R) - \int_{\p {\cal M}} \!\Tr(\Gamma \wedge \delta
\Gamma)~.}
The curvature two form was defined in \bm.

Next, we write the variation of the connection  in terms of the
underlying metric and simplify the resulting expression by
introducing Gaussian normal coordinates \ba. Starting from the
first term in \ed\ this procedure gives
\eqn\eab{ 2 \int_{\cal M}\! \Tr( \delta\Gamma \wedge R)=-
\int_{\cal M}\!d^3x \sqrt{g} \,\delta g_{\gamma\rho} (\nabla_\beta
R^{\beta\rho}_{~~\mu\nu}) \epsilon^{\gamma\mu\nu}+
2\int_{\partial{\cal M}} d^2 x\sqrt{g}~\delta g_{ij} R^{\eta
j}_{~~\eta k} \epsilon^{i\eta k}~, }
as detailed in Appendix B. The bulk term gives the correction to
the equations of motion. However, this term vanishes in the
important case of a space whose curvature is covariantly constant.
So, for example, the metrics of pure AdS$_3$ and BTZ black holes
are uncorrected by $S_{CS}(\Gamma)$.  The boundary term in \eab\
would seem to contribute to the boundary stress tensor; however,
the AdS$_3$ boundary is at large $\eta$, where the
Fefferman-Graham expansion \bi\ applies. As we detail in appendix
A, $R^{\eta j}_{~~\eta k}$ is proportional to $\delta^j_k$ in this
limit. This means the last term in \eab\ vanishes.

At this point we have shown that the first  term in \ed\ does not
contribute to the boundary stress tensor. Before evaluating the
second term in \ed\ explicitly, we can determine its large $\eta$
behavior  using the Fefferman-Graham expansion \bi\ and find,
confusingly, that it diverges at the boundary. Happily, the
explicit computation (in Appendix B) shows that the leading terms
for large $\eta$ cancel, leaving a finite answer. Due to this
cancellation, no boundary counterterms are needed beyond those
required by Einstein gravity (discussed in section 2.2). More
importantly, it shows that the total boundary variation of
$S_{CS}(\Gamma)$
\eqn\eb{ \delta S_{CS}(\Gamma) = {\rm bulk} - \int_{\p {\cal M}} \!\Tr(\Gamma \wedge \delta \Gamma)
 =  {\rm bulk}+ \half \int_{\p {\cal M}} \! d^2x
\sqrt{\go} T^{ij} \delta \go_{ij}~,}
comes purely from the leading asymptotic metric $\delta g^{(0)}$.
As discussed in section 3.1, this is a requirement for the existence
of a good variational principle.

The explicit result for the contribution of the  Chern-Simons term
to the stress tensor is of the form
\eqn\ec{ T^{ij} = t^{ij} + X^{ij}~,}
where
\eqn\ebb{ t^{ij}= {2 \over \ell^2} \left( \glt^{ik}
\epsilon^{lj} + \glt^{jk} \epsilon^{li} \right)\go_{kl}~, }
is the
contribution from the extrinsic  part of the connection, {\it
i.e.} the $\Gamma^\eta_{mn}$ and $\Gamma^m_{\eta n}$ components,
and $X^{ij}$ is defined through
 \eqn\ef{\half \int\! d^2x \sqrt{\go}  X^{ij}\delta g_{ij}^{(0)}= -
\int\! d^2x \sqrt{\go} ~ \!\Gamma^i_{jk} \delta \Gamma^j_{il}  {\epsilon^{kl}}
~,}
where the connection is formed from the boundary metric $\go_{ij}$.
Since $t^{ij}$ depends on $g^{(2)}_{ij}$, it is sensitive to the
precise geometry of the bulk space-time, rather than just its
conformal structure at infinity. We can think of this as a
dependence on the state of the theory. In contrast, $X^{ij}$ is
formed from $\go_{ij}$ alone, and defined so that it
vanishes when the connection constructed
from $\go_{ij}$ is trivial. In explicit computations with a fixed
conformal structure $X^{ij}$ appears as a background constant.
Indeed, in most examples, including the rotating BTZ black hole
in standard coordinates, and pure AdS$_3$ expressed in
global or Poincar\'{e} coordinates, we have $X^{ij}=0$, and
hence $T^{ij} = t^{ij}$. The explicit result for the stress
tensor in these geometries is given in the next subsection.

Let us also comment on the trace anomaly in the presence of
a Chern-Simons term. It is manifest from \ebb\ that the
trace of the state-dependent contribution vanishes.
Also, for a rigid (position independent) Weyl transformation
$\delta \go_{ij} = \delta\sigma \go_{ij}$ we have
$\delta\Gamma^j_{kl}=0$ and so $X^{ij}$ vanishes upon integration.
Thus, if we match the trace of the stress tensor to the usual covariant
form of the trace anomaly
\eqn\eg{ T^i_i = -{c_0 \over 12} R~,}
we find that $c_0$ is uncorrected by the Chern-Simons term.
In the present context the boundary theory is not diffeomorphism invariant so there
may be additional, non-covariant, terms on the right hand
side of \eg. However, such terms vanish upon integration.

\subsec{Examples: global AdS$_3$ and the BTZ black hole}
The metric of the rotating BTZ black hole is \btz\
\eqn\eh{ds^2 = -{(r^2-r_+^2)(r^2-r_-^2) \over \ell^2 r^2} dt^2 +
{\ell^2 r^2 \over (r^2-r_+^2)(r^2-r_-^2)} dr^2 +r^2(d\phi-{r_+ r_-
\over \ell r^2}dt)^2~.}
In this subsection we work in Lorentzian signature in order to
avoid awkward imaginary angular momenta. We define
\eqn\ei{ m = {r_+^2 + r_-^2 \over 8 G_3 \ell^2}~, \quad j = {2r_+
r_- \over 4 G_3 \ell}~.}
In this parametrization global AdS$_3$ is the special case $m=-{1
\over 8 G_3},~ j=0$, and AdS$_3$ in Poincar\'{e}  coordinates
corresponds to $m=j=0$.

In ordinary Einstein gravity $m$ and $j$ are  identified with the
mass and angular momentum of the black hole. We wish to see how
this is modified due to the Chern-Simons term
\eqn\eia{ S_{CS}(\Gamma) = \beta \int_{\cal M}
\!\Omega_3(\Gamma)~.}
Transforming \eh\ into Gaussian normal coordinates using
\eqn\eja{(r/\ell)^2 =   e^{2\eta/\ell}+ 4G_3m+\cdots~,}
and expanding for large $\eta$ we have
\eqn\ej{ ds^2 = d\eta^2 + e^{2\eta/\ell}( -dt^2+\ell^2 d\phi^2) +
(4G_3m dt^2+ 4G_3m\ell^2 d\phi^2 -8G_3jdt d\phi) + \ldots~,}
The expressions in brackets in this equation are  identified with
the components $\go_{ij}$ and $g^{(2)}_{ij}$ of the
Fefferman-Graham expansion \bi. The stress tensor, including the
contribution \bj\ from the Einstein-Hilbert term, is
\eqn\ek{ T_{ij} = {1 \over 8 \pi G_3 \ell} \left[ g^{(2)}_{ij} - g^{(0)}_{ij} g^{(2)}_{kl}g_{(0)}^{kl}  \right]+{2\beta
\over \ell^2} \left[g^{(2)}_{ik}\epsilon_{lj} g_{(0)}^{kl}
+ g^{(2)}_{jk} \epsilon_{li} g_{(0)}^{kl}\right]~.}
Inserting $g^{(0)}$ and $g^{(2)}$ from \ej, and taking the orientation $\epsilon_{t\phi} = -\ell$,
we find
\eqn\el{\eqalign{ T_{tt}& = {m \over 2\pi \ell} -{16 \beta G_3
j\over \ell^3}~, \cr T_{\phi \phi} & = {m \ell \over 2\pi}-{16 \beta
G_3 j \over \ell}~, \cr T_{t\phi} &  =-{j\over 2\pi \ell} +{16 G_3
\beta m \over \ell}~. }}
Therefore, the corrected formulas for the mass and angular
momentum of the BTZ black hole are
\eqn\em{\eqalign{  M & = 2\pi \ell T_{tt} = m -{32 \pi \beta G_3 j
\over \ell^2}~, \cr J& = -2\pi \ell T_{t\phi} = j - 32 \pi \beta G_3
m~. }}
Alternatively, we can express these results in terms of the Virasoro
generators as
\eqn\en{\eqalign{ L_0 - {c_L \over 24} & = {M\ell-J \over
2}=\left(1+{32\pi \beta G_3 \over\ell}\right) {m\ell-j \over 2} \cr
\tilde{L}_0 - {c_R \over 24} & = {M\ell+J \over 2} =\left(1-{32\pi
\beta G_3 \over\ell}\right) {m\ell+j \over 2}~. }}

An important special case is global AdS$_3$ where $m=-{1
\over 8 G_3}$ and $j=0$ and so
\eqn\eo{M_{AdS_3} = -{1 \over 8 G_3}, \quad J_{AdS_3} = 4\pi
\beta~.}
In particular, global AdS$_3$ carries a non-vanishing angular
momentum. For the dual CFT perspective on this result, recall that
global AdS$_3$ corresponds to the NS-NS vacuum which has $L_0 =
\tilde{L}_0=0$, and thus
\eqn\ep{c_L= {3 \ell \over 2G} + 48 \pi \beta~, \quad c_R
={3\ell \over 2G} - 48\pi \beta~,}
which gives the correction to the usual Brown-Henneaux result for
the central charge. With this in hand, we can reexpress \eo\ as
\eqn\eq{M_{AdS_3}\ell =-{c_L+  c_R \over 24}~,
\quad J_{AdS_3} ={c_L-c_R \over 24} ~.}
This shows that the ground state angular momentum is due to the
asymmetry of the central charges, which implies a non-cancellation
between the left and right moving zero point momenta.

\newsec{Application to black hole entropy in string theory}

In \us, building on arguments in \HMM, we showed that
gravitational anomalies can be used to compute higher derivative
corrections to black hole entropy.  This explains the success of
recent computations which  take into account a subset of higher
derivative terms and find agreement  with microscopic entropy
counting. As explained in \us, these terms include the effects of
anomalies, and this, along with symmetries, is enough to determine
the corrections to the entropy.  In this section we review this
argument.

Our argument applies to black holes which have a BTZ factor.  To
make contact with other papers in the literature which deal with
near horizon AdS$_2$ geometries, we note that if we Kaluza-Klein
reduce a BTZ black hole along the horizon direction we recover a
near-horizon AdS$_2$ factor \StromingerYG.   However, this
reduction obscures some of the symmetries, namely the existence of
left and right moving Virasoro algebras, and so we will stick to
the BTZ description.

In \us\ we showed that in a general higher derivative theory of
gravity the entropy of a BTZ black hole is
\eqn\fa{ S = 2\pi \left[ \sqrt{{c_L h_L \over 6}} + \sqrt{{c_R h_R
\over 6}} ~\right]~,}
with
\eqn\fb{ h_L = {M\ell -J \over 2}, \quad h_R = {M\ell +J \over
2}~.}
In a diffeomorphism invariant theory \fa\ is equivalent to Wald's
generalized entropy formula \wald, as discussed in
\refs{\saidasoda,\us}.  \fa\ is actually  more general (in the
case of BTZ black holes) in that it also applies in the presence
of Chern-Simons terms.  \fa\ is just the leading saddle point
contribution to the entropy.  More generally, one has to perform
an inverse Laplace transform to convert a partition function to a
density of states,  but we omit this here; for more details see
the discussion in the last section of \us.

So to compute black hole entropy  we just need to determine the
central charges $c_{L,R}$.  In a general theory this requires
knowing the full action and carrying out the c-extremization
procedure discussed in \us.  But in string theory examples with
enough supersymmetry, the central charges are related to
anomalies, and these can be computed exactly purely from knowledge
of Chern-Simons terms.

A case of particular interest corresponds to wrapping M5-branes on
a 4-cycle $P_0$ of a Calabi-Yau threefold.   This gives rise to a
magnetic string in $D=5$, described at low energies by a CFT with
$(4,0)$ supersymmetry.   In this CFT, the gravitational anomaly
determines $c_L-c_R$, and the anomaly with respect to the
leftmoving $SU(2)$ R-symmetry (which always exists as part of the
superconformal algebra) determines $c_L$.  For us, the key point
is that from a supergravity point of view these anomalies are
determined from Chern-Simons terms.  Furthermore, in string theory
the coefficients in front of Chern-Simons terms are known exactly,
since they are required for anomaly cancellation.

For the $D=5$ case just mentioned, the relevant Chern-Simons term
is
\eqn\fc{ S   =\half\left({1 \over 2\pi}\right)^2 {c_2 \cdot P_0
\over 48} \int F^{(2)} \wedge \Omega_3~, }
where $c_2\cdot P_0$ denotes the second Chern class of the
$P_0\subset CY_3$, and $F^{(2)}$ has flux corresponding to charge
$q$, as in \df. The magnetic string supports an AdS$_3 \times S^2$
solution\foot{The scalars are fixed by the 5D attractor mechanism
\refs{\FerraraDD\ChamseddinePI-\KLatt}.}, so we can follow the
analysis leading to \di, with
\eqn\fd{ \beta =-\half\left({1 \over 2\pi}\right) {c_2 \cdot q
\over 48} ~.}
From \de\ we learn that
\eqn\fe{ c_L -c_R = 96\pi \beta = -\half c_2 \cdot q~.}
The leftmoving supersymmetry implies that the $SU(2)$ current
algebra has level $k=c_L/6$, and associated anomaly
\eqn\fb{ \delta S = -{c_L \over 96 \pi} \int \! \Tr(\Lambda dA)~.}
Comparing with \dj\ we find
\eqn\fc{ \Delta c_L = 96 \pi \beta = \half c_2 \cdot q~.}
  Here we have indicated that \fc\ just gives the correction to
$c_L$; there is also a term coming from the Chern-Simons term in
the lowest order supergravity theory. Altogether one gets
\eqn\fd{ c_L = C_{IJK}q^I q^J q^K +
 \half c_2 \cdot q, \quad c_R C_{IJK}q^I q^J q^K +  c_2 \cdot q~.}
As we have stressed, this result is exact.  Any correction would
imply a noncancellation of anomalies in the theory of an M5-brane
in M-theory, and so would lead to an inconsistency.

Inserting this result into \fa\ we get a result for black hole
entropy including higher order derivative corrections.  This
result is in agreement with the much more involved derivation in
\CardosoFP, which requires knowing the full supergravity action at
the level of $R^2$ terms in $D=4$.  The only assumptions we needed
to make was that there exists an AdS$_3 \times S^2$ vacuum
corresponding to our microscopic system, and that the effective
supergravity theory respects the $(4,0)$ symmetry.   If the
symmetry assumption is false it will represent a breakdown of the
AdS/CFT correspondence.

In other approaches it is unclear why even higher derivative
corrections, of $R^4$ type and beyond, don't destroy the
agreement. Such terms are certainly present, and do individually
lead to unwanted corrections.  The point is that the $(4,0)$
symmetry implies that the sum of all such contributions must
vanish.  Any non-vanishing total result would, by symmetry, lead
to a shift in the anomalies, and this would lead to an
inconsistency as we have discussed.

We emphasize that while we have used supersymmetry of the
underlying theory, we have not demanded that our black hole
preserves supersymmetry.  Thus the entropy formula \fa\ applies
even to non-supersymmetric BTZ black holes.  These correspond to
the case when $h_L$ is nonvanishing.

\bigskip
\noindent {\bf Acknowledgments:} \medskip \noindent We thank David
Kutasov and Arvind Rajaraman  for discussions. FL thanks the Aspen
Center for Physics for hospitality during the completion of this
work. The work of PK is supported in part by the NSF grant
PHY-00-99590. The work of FL is supported  by DoE under grant
DE-FG02-95ER40899.

\appendix{A}{Gaussian normal coordinates and the
Fefferman-Graham expansion}
In this appendix we collect a number of useful formulae pertaining
to the use of Gaussian normal coordinates and the Fefferman-Graham
expansion. The formulae are valid in arbitrary dimension.

We define the Gaussian normal coordinates by the line element
\eqn\baa{ ds^2 = d\eta^2 + g_{ij} dx^i dx^j~. }
In these coordinates the extrinsic curvature of a surface at fixed
$\eta$ is given by $K_{ij} = \half\partial_\eta g_{ij}$. The
nonvanishing Christoffel symbols are
\eqn\bab{
\Gamma^\eta_{ij} = -K_{ij}~,~~\Gamma^i_{\eta j}= K^i_j~,
}
as well as the $\Gamma^k_{ij}$ which is the same from the bulk and the boundary
point of view.

The components of the Riemann tensor are
\eqn\bac{\eqalign{ R^i_{~jk\eta} &= \d\nabla^i K_{kj}-\d\nabla_j
K^i_k ~,\cr R^i_{~\eta j\eta} &= -K^k_j K^i_k - \partial_\eta K^i_j
~,\cr R^i_{~jkl} &= \d R^i_{~jkl} +K_{kj}K^i_l - K_{lj}K^i_k~. }}
By contraction, we find the Ricci tensor
\eqn\bad{\eqalign{ R_{\eta\eta} & = -\Tr K^2 -\p_\eta \Tr K~, \cr
R_{\eta i} &=  \d\nabla^j K_{ij} -\d\nabla_i \Tr K~,  \cr R^i_{~j}&\d \!R^i_{~j} - K^i_{~j} \Tr K - \p_\eta K^i_{~j}~, } }
and the Ricci scalar
\eqn\bae{ R = \d \!R -(\Tr K)^2- \Tr K^2 -2 \p_\eta \Tr K~.}

Spaces that are asymptotically AdS$_{d+1}$ allow the Fefferman-Graham
expansion\foot{We neglect a possible term that is linear in $\eta$; it does not
play any role in our considerations.}
\eqn\baf{g_{ij}  = e^{2\eta/\ell}g^{(0)}_{ij} + g^{(2)}_{ij} + \ldots~.}
The corresponding inverse expansion is
\eqn\bag{g^{ij}  = e^{-2\eta/\ell}g_{(0)}^{ij} -
e^{-4\eta/\ell}g_{(2)}^{ij} + \ldots~,}
with the understanding that indices on $g_{(2)}$ are raised and lowered
using $g_{(0)}$. Some useful formulae for the expansion of the
extrinsic curvature are
\eqn\bah{\eqalign{
K_{ij} & = {1\over\ell} e^{2\eta/\ell} g_{ij}^{(0)} + 0 + \cdots~, \cr
K^i_j &= {1\over\ell} \delta^i_j -{1\over\ell} e^{-2\eta/\ell} g^{ij}_{(2)} g^{(0)}_{lj}+\cdots~.
}}
The expansions for the Christoffel symbols with one extrinsic index
follows from \bab. The leading term in the symbol $\Gamma^k_{ij}$ is
simply formed from $g_{(0)}$  and is $\eta$ independent.

The components of the Riemann curvature expand as
\eqn\bai{\eqalign{
R^i_{\eta k\eta} &= -{1\over\ell^2}\delta^i_k + 0 + \cdots ~,\cr
R^i_{jk\eta} &= -{1\over\ell} e^{-2\eta/\ell} \left[ g_{jm}^{(0)} ~\d\nabla^i g^{ml}_{(2)}-
\d\nabla_j g^{il}_{(2)} \right]g^{(0)}_{lk}+\cdots~,\cr
R^i_{jkl}&= {1\over\ell^2} e^{2\eta/\ell} \left( g^{(0)}_{kj} \delta^i_l - g^{(0)}_{lj} \delta^i_k \right)
+ \left[ \d R^i_{jkl} -{1\over\ell^2} \left( g^{(0)}_{kj} g^{im}_{(2)} g^{(0)}_{ml}
- k\leftrightarrow l \right) \right]+\cdots~.
}}

\appendix{B}{Explicit variations of the Chern-Simons Term}
In this appendix we give some details on the computation of the contribution to
the boundary stress tensor from the Chern-Simons term \ea. The variation with
respect to the metric is
\eqn\aba{
\delta S_{CS}(\Gamma) = 2\int_{\cal M} {\rm Tr} (\delta\Gamma\wedge R)
- \int_{\partial{\cal M}} {\rm Tr} (\Gamma\wedge \delta\Gamma)~.
}
In components, the first term reads
\eqn\aaa{\eqalign{
{\rm Tr} (\delta\Gamma \wedge R) & = \delta \Gamma^\alpha_{~\beta} \wedge R^\beta_{~\alpha}\cr
&= \delta\Gamma^\alpha_{~\beta\gamma} ~\half R^\beta_{~\alpha\mu\nu} dx^\gamma
\wedge dx^\mu\wedge dx^\mu \cr
& = \half g^{\alpha\rho}\left( \nabla_\gamma\delta g_{\beta\rho}
+\nabla_\beta\delta g_{\gamma\rho} -\nabla_\rho\delta g_{\beta\gamma}\right)
\half R^\beta_{~\alpha\mu\nu} \epsilon^{\gamma\mu\nu} \sqrt{g} d^3 x\cr
&= \half (\nabla_\beta\delta g_{\gamma\rho}) R^{\beta\rho}_{~~\mu\nu}
\epsilon^{\gamma\mu\nu} \sqrt{g} d^3 x~.
}}
After integration by parts we thus find a bulk term of the form
\eqn\aab{
\delta S_{CS}(\Gamma) =- \int_{\cal M}\delta g_{\gamma\rho} (\nabla_\beta R^{\beta\rho}_{~~\mu\nu}) \epsilon^{\gamma\mu\nu} + \cdots ~,
}
which contributes to the bulk Einstein equation. This term vanishes
in situations where the curvature is covariantly constant, such  as
in AdS$_3$ or BTZ.  Since we're only interested in the stress
tensor, the bulk term \aab\ will play no further role. The boundary
term arising from the partial integration takes the form
\eqn\aac{ \delta S_{CS}(\Gamma) =\int_{\partial{\cal M}} d^2
x\sqrt{g}~\delta g_{ij} R^{\eta j}_{~~\mu\nu} \epsilon^{i\mu\nu} =
2\int_{\partial{\cal M}} d^2 x\sqrt{g}~\delta g_{ij} R^{\eta
j}_{~~\eta k} \epsilon^{i\eta k} =0~. }
This term vanishes identically because, according to \bai, we have $R^{\eta j}_{~~\eta k}\propto \delta^j_k$
up to terms that vanish as the boundary is taken to infinity.

The second term in \aba\ is explicitly a boundary term. We begin
by expanding in intrinsic and extrinsic quantities as
\eqn\aac{\eqalign{
\delta S_{CS}(\Gamma) &= - \int_{\partial{\cal M}} \Gamma^\alpha_{\beta i} \delta
\Gamma^\beta_{\alpha j} dx^i\wedge dx^j  \cr
& =  - \int_{\partial{\cal M}} \left[ \Gamma^\eta_{ki}\delta\Gamma^k_{\eta j}
+ \Gamma^k_{\eta i}\delta\Gamma^\eta_{kj} + \Gamma^k_{li} \delta\Gamma^l_{kj} \right] \epsilon^{ij}
\sqrt{g} d^2 x \cr
& =  \int_{\partial{\cal M}} \left[ 2K^k_i\delta K_{kj} -  \Gamma^k_{li} \delta\Gamma^l_{kj} \right] \epsilon^{ij}
\sqrt{g} d^2 x~.
}}
Using \bah\ we find the Fefferman-Graham expansion of the first term
\eqn\aad{
K^i_k\delta K_{ij} \epsilon^{kj}
= -{1\over\ell^2} g^{il}_{(2)} g^{(0)}_{lk} \delta g^{(0)}_{ij} \epsilon^{kj}+\cdots ~,
}
which amounts to the contribution
\eqn\aae{
t^{ij} = {2\over\ell^2} \left[ g^{il}_{(2)} g^{(0)}_{lk} \epsilon^{jk} +i\leftrightarrow j \right]~,
}
to the stress tensor. This contribution depends  on $g_{(2)}$ and
so on the state. The second term in \aac\ is finite for large
$\eta$ and, in the limit, depends only on $g_{(0)}$. Moreover, the
variation considered in this appendix was general and we have seen
explicitly that no $\delta g_{(2)}$ dependence remains for large
$\eta$.

\listrefs
\end